\documentclass{optica-article}

\journal{opticajournal} 

\articletype{Research Article}

\usepackage{lineno}

\begin{document}
\title{Accidental coincidences in camera-based high-dimensional entanglement certification}

\author{Raphael Guitter \authormark{1,*}, Baptiste Courme \authormark{1,2}, Chloé Vernière \authormark{1}, Peter Svihra 
\authormark{3}, Andrei Nomerotski 
\authormark{3} and Hugo Defienne \authormark{1}}

\address{\authormark{1}Sorbonne Université, CNRS, Institut des NanoSciences de Paris, INSP, F-75005 Paris, France\\
\authormark{2}Laboratoire Kastler Brossel, ENS-Universite PSL, CNRS, Sorbonne Universite, College de France, 24 rue Lhomond, 75005 Paris, France\\}

\address{\authormark{3} Faculty of Nuclear Sciences and Physical Engineering,Czech Technical University, 160 00 Prague, Czech Republic}
\email{\authormark{*}raphael.guitter@insp.jussieu.fr}

\begin{abstract*} 
High-dimensional entangled states, such as spatially-entangled photon pairs produced by Spontaneous Parametric Down-Conversion (SPDC), are a key resource for quantum technologies. In recent years, camera-based coincidence counting approaches have considerably improved the ability to characterize them in terms of speed and dimensionality. However, these methods have limitations, including in most of them the necessity to subtract accidental coincidences. Here, we study the role of these accidentals in entanglement certification for a single-photon avalanche diode (SPAD) array and an intensified time-stamping (Tpx3Cam) camera. Using both Einstein-Podolsky-Rosen (EPR) and entropy-based criteria, we show that the level of accidental coincidences - determined by the temporal characteristics of the camera - and whether they are subtracted critically impact entanglement certification.
In particular, we demonstrate that current single-photon camera technologies enable entanglement certification without accidental subtraction only if a Gaussian approximation is applied to the measured two-photon state.
Our work is important for developing quantum-optics application in adversarial scenarios, such as high-dimensional quantum key distribution (HD-QKD), and also for loophole-free experimental testing of quantum foundations.
\end{abstract*}

\section{Introduction}
As discussed by Lantz et al.~\cite{PhysRevA.110.023701}, the subtraction of accidental coincidences in quantum optics - a process equivalent to measuring the covariance rather than directly the second-order correlation function $G^{(2)}$ - has significant implications for the interpretation of experimental results. For instance, it can introduce loopholes in experiments testing the foundations of quantum physics, such as Bell inequality violations~\cite{larsson_loopholes_2014}, or security flaws in applications such as quantum key distribution (QKD)~\cite{liu_experimental_2013}. 
Thanks to the high performance of modern single-pixel single-photon detectors, such as Superconducting Nanowire Single-Photon Detector (SNSPDs)~\cite{natarajan_superconducting_2012}, subtracting accidental coincidences is no longer critical in experiments involving low-dimensional quantum states i.e. states containing only a few photons distributed across a small number of modes. However, as the dimension of the Hilbert space increases, and with it, the number of modes to be detected, it becomes necessary to use detector arrays or single-photon sensitive cameras. These emerging multi-pixel detection systems are less ideal than single-pixel detectors, and in such cases, subtracting accidental coincidences, or making assumptions about the detected state, often becomes necessary.

A prime example of quantum states entangled in high dimensions are photon pairs generated via Spontaneous Parametric Down-Conversion (SPDC), which naturally exhibit entanglement in the high-dimensional Hilbert space of transverse spatial modes~\cite{SPDC-Calcul-Phi}. These states are at the heart of correlation-based quantum imaging approaches~\cite{defienne_advances_2024}, but are also used for fundamental test of quantum mechanics~\cite{PhysRevLett.92.210403} and to develop high-dimensional quantum communication and computing approaches~\cite{mirhosseini_high-dimensional_2015,scarfe_spatial-mode_2025,brandt_high-dimensional_2020}. 
When the dimensionality of these states becomes large (e.g. exceeding 1000 modes), conventional single-outcome projective measurements - where all modes are sequentially scanned and projected onto a single-pixel detectors - become not only impractical due to acquisition time constraints, but also risky for certain applications because they leave the fair-sampling loophole open. Over the past ten years, single-photon-sensitive cameras have thus emerged as the tool of choice to detect and characterize them~\cite{Moreau_2019,madonini_single_2021,kundu_adaptively_2024, nomerotski_intensified_2022}. 

In particular, these cameras have been used to certify high-dimensional spatial entanglement, a crucial step toward the development of high-dimensional quantum information processing protocols. Using a practical Einstein-Podolsky-Rosen (EPR) criterion introduced by Reid et al.~\cite{reid_demonstration_1989,Cavalcanti_experimental_2009,reid_colloquium_2009} (EPR-Reid), entanglement certification has been explored with Electron-Multiplying Charge-Coupled Device (EMCCD) cameras~\cite{Edgar2012,PhysRevA.86.010101,lantz_einstein-podolsky-rosen_2015}, intensified Complementary Metal-Oxide-Semiconductor (iCMOS) cameras~\cite{Dąbrowski:17,dabrowski_certification_2018}, Single-Photon Avalanche Diode (SPAD) arrays~\cite{HugoCertif,eckmann_characterization_2020} and intensified time-stamping cameras such as the Tpx3Cam~\cite{Courme:23,li_rapid_2025}. However, none of the above-cited works have achieved violation of the EPR criterion without either subtracting accidentals or making assumptions about the detected quantum state.  Due to their limited temporal resolution, demonstrations with EMCCD cameras~\cite{Edgar2012,PhysRevA.86.010101,lantz_einstein-podolsky-rosen_2015} require both the subtraction of accidental coincidences and the modeling of the detected two-photon state, typically using a double-Gaussian model~\cite{fedorov_gaussian_2009}. The same assumptions have been used in works involving iCMOS and SPAD cameras~\cite{HugoCertif,eckmann_characterization_2020,Dąbrowski:17,dabrowski_certification_2018}, sometimes with additional cross-talk and artefacts corrections. Recent results obtained using the Tpx3Cam camera~\cite{Courme:23,li_rapid_2025} show that subtracting accidentals can be avoided; however, Gaussian state modeling is still required. 

In Refs.\cite{HugoCertif,li_rapid_2025}, the authors also use an alternative criterion to certify entanglement, derived from Refs.~\cite{Bavaresco2018,erker_quantifying_2017}. It estimates a lower bound to the entanglement dimensionality by detecting photons in two discrete mutually unbiased bases (MUBs). However, in these experimental demonstrations, photons are detected using subsets of pixels chosen from the continuous position and momentum bases (i.e. coarse-grained measurements), which do not strictly form two discrete MUBs~\cite{tasca_mutual_2018}. As a result, the criterion is applied under the assumption that the bases are mutually unbiased, yet the impact of this approximation on the estimated lower bound is not quantified and may be non-negligible. In contrast, the EPR-Reid criterion directly uses measurements in continuous position and momentum bases, making it a more reliable choice for our experimental configuration. 

Furthermore, the EPR-Reid criterion corresponds to a limiting case of a more general entropic entanglement criterion derived in Refs.~\cite{PhysRevLett.106.130402, PhysRevLett.103.160505}, which has been experimentally used in Refs.~\cite{PhysRevLett.106.130402, Schneeloch2019}, though without the use of a camera. This entropic approach can certify entanglement even in cases where the EPR-Reid criterion fails, while also avoiding the approximations inherent to discrete MUB-based criteria.

In our work, we study the influence of accidental coincidences on the certification of high-dimensional entanglement using the EPR-Reid and entropic criteria with two types of cameras: a SPAD camera from Micro Photon Devices (model Hermes) and a Tpx3Cam (model Phoebe). By varying the coincidence window duration, we demonstrate that the rate of accidental coincidences plays a critical role in the violation of the criteria, preventing conclusive certification once the coincidence window exceeds a few hundred nanoseconds. Finally, we show that even these state-of-the-art cameras do not allow a strict violation of neither the EPR-Reid criterion as originally defined nor the entropic criterion, instead requiring a Gaussian model for the detected two-photon state.

\section{Experimental Setup}
\begin{figure}[htbp]
\centering\includegraphics[width=12cm]{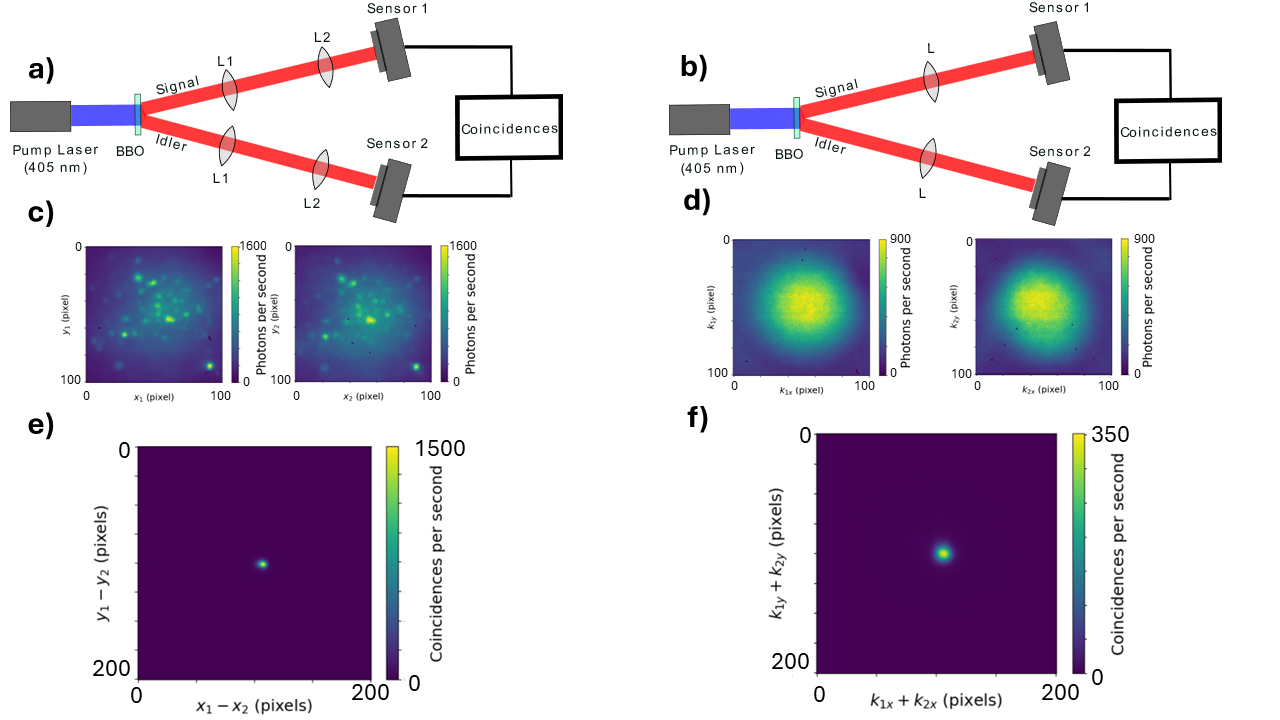}
\caption{a),b) Simplified experimental setups for correlation measurement. A continuous-wavelength collimated 405nm laser diode illuminates a $\beta$-baryum borate (BBO) nonlinear crystal to produce the spatially entangled photon pairs at 810nm by Spontaneous Parametric Down-Conversion (SPDC). The signal and idler photons of each pair are then imaged with a telescope (a) or Fourier-imaged with a single lens (b) into a single photon sensitive camera (SPAD array or Tpx3Cam). In reality a single camera is used in each experiment, the two sensors corresponding to two separate areas of the camera. In the configuration using the SPAD camera, a type-I BBO crystal is employed, followed by a non-polarizing beam splitter to separate the photon pairs into two arms. In contrast, the configuration with the Tpx3Cam camera uses a type-II BBO crystal of the same thickness, followed by a polarizing beam splitter for photon separation. Full experimental setups are presented in Supplementary document section 1. c),d) Intensity images obtained with the Tpx3Cam in the imaging (c) and Fourier imaging (d) configurations. The bright dots in c) come from dust on the crystal surface that is imaged onto the camera. e) Minus-coordinate projection of the coincidences of events detected in c), that highlights strong correlations in position in the photon pairs. f) Sum-coordinate projection of of the coincidences of events detected in d), that highlights strong anti-correlations in momentum.}
\label{fig:Setup}
\end{figure}

Spatially-entangled photon pairs are produced via Spontaneous Parametric Down-Conversion (SPDC) by illuminating a $\beta$-barium borate (BBO) nonlinear crystal with a continuous-wave 405 nm laser.
The signal and idler photons at 810nm are each sent through a different arm onto a single-photon sensitive camera. 
As shown in Figures~\ref{fig:Setup}a and b, each arm can take two different configurations. In the imaging configuration (Fig.~\ref{fig:Setup}a), the plane of the crystal is imaged onto the camera by two lenses in a 4f telescope configuration. Positions on the camera thus correspond to photon positions in the crystal plane i.e. the camera performs measurements in the position basis. In the Fourier imaging configuration (Fig.~\ref{fig:Setup}b), a single lens in a 2f configuration performs an optical Fourier transform between the plane of the crystal and this of the camera. Positions on the camera then correspond to photon momenta in the crystal plane i.e. the camera performs measurements in the momentum basis. Figures~\ref{fig:Setup}c and d show examples of intensity images measured by the Tpx3Cam camera at the output in both configurations.  

To certify entanglement, coincidence measurements between the idler and signal photons must be performed in each of these bases. In our work, we are using two different cameras : SPAD and Tpx3Cam. The first camera is an array of $64 \times 32$ single-photon avalanche diodes that each correspond to a pixel~\cite{madonini_single_2021}. This camera is frame-based, capturing successive images with adjustable exposure times  $\Delta T$ ranging from 10 ns to 10 $\mu$s. Every pair of photons appearing on the same frame is then recorded as a coincidence. The second camera is a Tpx3Cam~\cite{FisherLevine2016,NOMEROTSKI201926,Tpx3Cam,Vidyapin2023}. 
It is an event-based camera: instead of capturing successive frames, it records individual photon detection events $i$, including their position ($x_i,y_j$) within the $256 \times 256$ pixel array and their detection time $t_i$, with an effective time resolution of $6$ ns. Coincidences are extracted during post-processing by selecting all pairs of events $(i,j)$ that satisfy $|t_i - t_j| < \Delta T$, where $\Delta T$ is a chosen coincidence time window. 
These multi-pixel coincidence detections allow to compute the spatial joint probability distribution (JPD) which corresponds to the probability of detecting simultaneously a signal photon at position $(x_1,y_1)$ and an idler photon at position $(x_2, y_2)$. Formally, it is a direct measurement of the second-order correlation function between the two pixels integrated over the coincidence time window.

To certify entanglement, we compute both the EPR-Reid criterion~\cite{reid_colloquium_2009} and the entropic criterion~\cite{PhysRevLett.103.160505} from correlation measurements performed in the two bases.
If the detected state is separable, the EPR-Reid criterion states that  
\begin{equation}
    \Delta x_- \Delta k_+ \geq \frac{1}{2},
    \label{inequality}
\end{equation}  
where $\Delta x_-$ and $\Delta k_+$ are the position and momentum correlation width for a given spatial transverse axis ($\vec{x}$ or $\vec{y}$), respectively. $\Delta x_-$ and $\Delta k_+$ correspond to the widths of the peaks observed in the JPD projections along the minus-coordinate axis $(x_1 - x_2, y_1 - y_2)$ and the sum-coordinate axis $(x_1 + x_2, y_1 + y_2)$, respectively, such as those shown in Figures~\ref{fig:Setup}e and f. As in previous works~\cite{Edgar2012,PhysRevA.86.010101,lantz_einstein-podolsky-rosen_2015,Dąbrowski:17,dabrowski_certification_2018,HugoCertif,eckmann_characterization_2020,Courme:23,li_rapid_2025}, these quantities can be estimated by performing Gaussian fits, based on the assumption that the two-photon state generated by the crystal is pure and described by a Double-Gaussian wavefunction~\cite{fedorov_gaussian_2009}. 

\begin{figure}[htbp]
\centering\includegraphics[width=12cm]{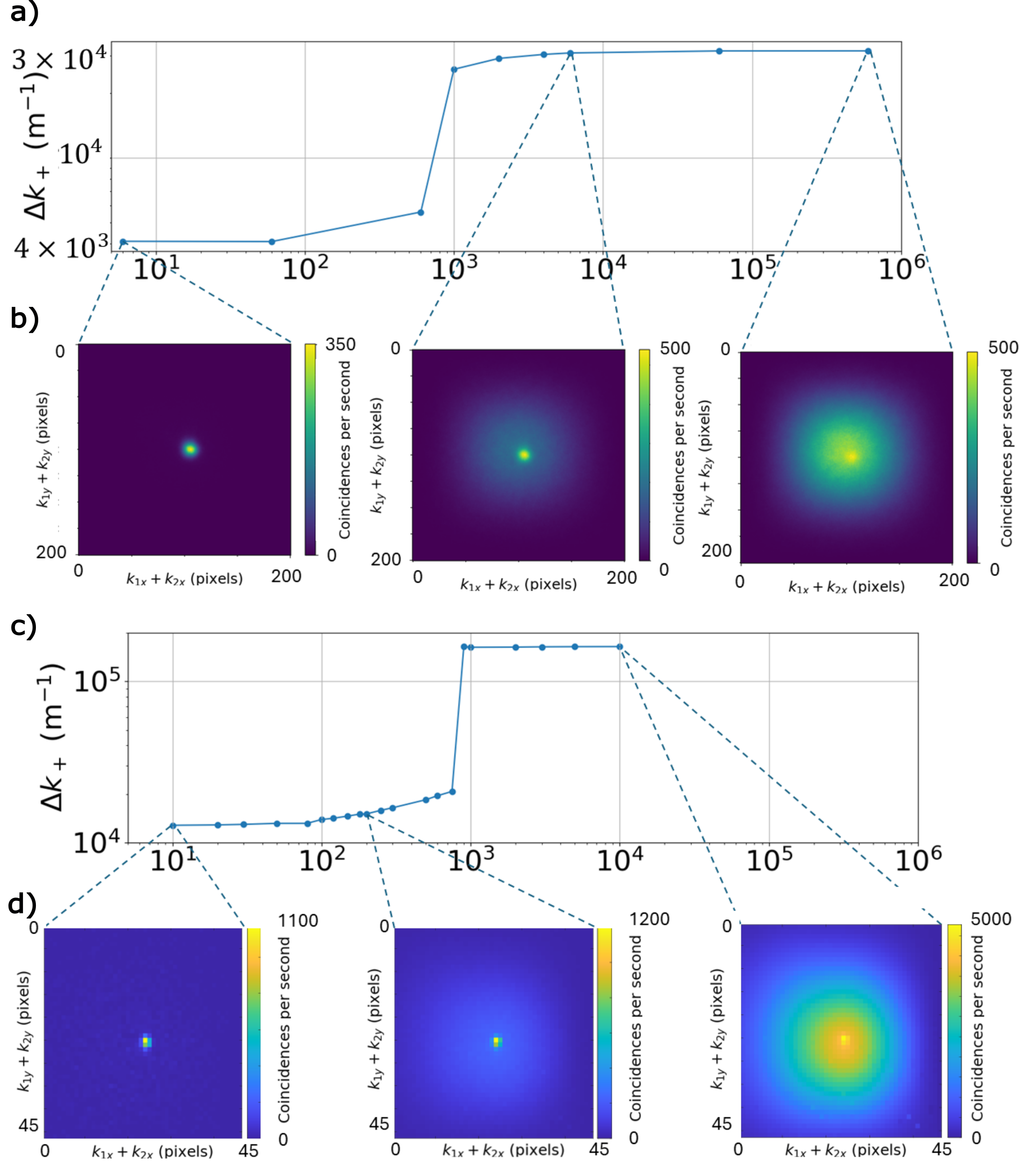}
\caption{a) Evolution of the momentum correlation width $\Delta k_+$ with the time window $\Delta T$ using the Tpx3Cam camera. b) JPD sum-coordinate projections obtained for $\Delta T = 6$ ns (left), $\Delta T = 4\mu$s (middle) and $\Delta T = 600\mu$s (right). c)  Evolution of the momentum correlation width $\Delta k_+$ with $\Delta T$ using the SPAD camera. d) JPD sum-coordinate projections obtained for $\Delta T = 10$ ns (left), $\Delta T = 500$ns (middle) and $\Delta T = 10\mu$s (right).
In both cases, for $\Delta T$ of a few nanoseconds - well below the average time separation of approximately 100 ns between photon pairs in our setup - only genuine coincidences are recorded, keeping the correlation peak narrow and $\Delta k_+$ small. As $\Delta T$ increases, accidental coincidences dominate, broadening the peak and increasing $\Delta k_+$ by an order of magnitude. Each acquisition lasted 8 seconds for the Tpx3Cam and 650 seconds for the SPAD camera.}
\label{fig:Deltak}
\end{figure}

Figures~\ref{fig:Deltak} show the values of $\Delta k_+$ for both the Tpx3Cam (a and b) and the SPAD camera (c and d), as a function of $\Delta T$ (see supplementary document for $\Delta x_-$). 

In addition to the narrow correlation peak observed on both cameras at low $\Delta T$, a broader background emerges with increasing $\Delta T$, leading to larger measured $\Delta k_+$ values (Figs.~\ref{fig:Deltak}b and d). 
This background arises from accidental coincidences, originating from the detection of two photons from different pairs or from a photon from a pair and a noise event (e.g. dark noise or stray light).
As shown in Figures~\ref{fig:Deltak}a and c, it becomes dominant for $\Delta T \gtrsim 800$ ns. This timescale corresponds to the average interval between two photon pairs emitted by the source, which therefore operates at a pair rate of approximately $10^7$ pairs/s.

\begin{figure}[htbp]
 \centering\includegraphics[width=12cm]{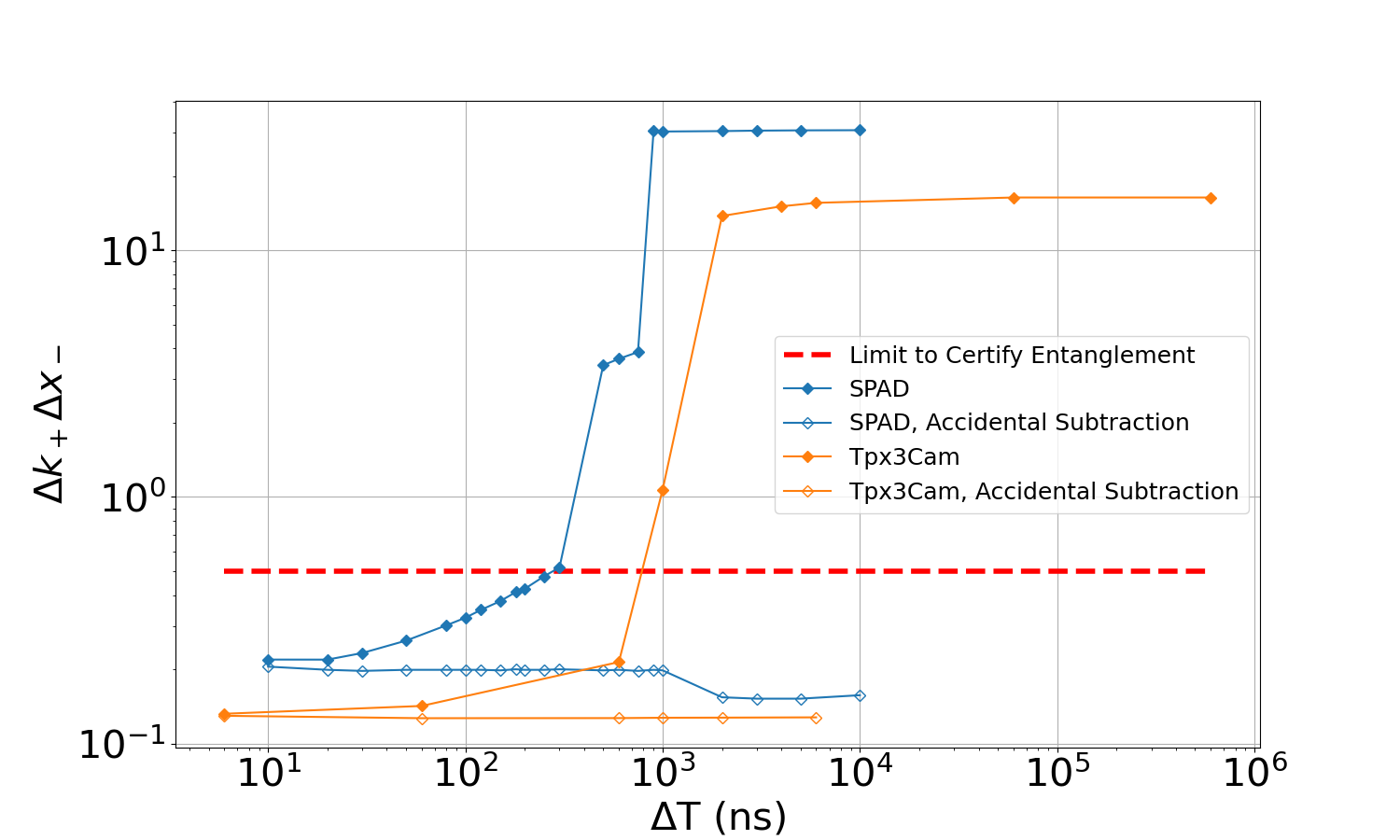}
 \caption{EPR-Reid factor $\Delta k_+ \Delta x_-$ as a function of the time resolution $\Delta T$ of the coincidence counting apparatus for both cameras, obtained from the raw coincidence counts and from the same counts after accidental coincidence subtraction. A value lower than $1/2$ (red line) allows to certify entanglement. For the raw data, it is achieved for time resolutions lower than a few hundreds of nanoseconds. Both correlation widths are estimated by Gaussian fitting of the correlation peak. When the accidental coincidences are subtracted in post processing, entanglement can be certified for all values of $\Delta T$. Each acquisition lasted 8 seconds for the Tpx3Cam and 650 seconds for the SPAD camera.}
 \label{fig:EPR_Delta}
 \end{figure}

This widening of both correlation peaks translates into the EPR-Reid factor $\Delta x_- \Delta k_+$, as shown in Figure~\ref{fig:EPR_Delta}. For small enough values of $\Delta T$, below 150 ns for the SPAD array and 600 ns for the Tpx3Cam, the measured EPR-Reid factor falls below 1/2, allowing one to certify entanglement (Equation~\eqref{inequality}). 
We measure minimum values of $\Delta x_- \Delta k_+ = 0.219 \pm 0.005$ at $\Delta T = 10$ ns with the SPAD camera, and $\Delta x_- \Delta k_+ = 0.132 \pm 0.001$ at $\Delta T = 6$ ns with the Tpx3Cam. The higher value obtained with the SPAD camera is attributed to its lower spatial resolution ($32 \times 64$ pixels), which leads to an overestimation of peak widths in Gaussian fits. These measurements are consistent across both $\vec{x}$ and $\vec{y}$ axes, confirming entanglement in both spatial directions.
When $\Delta T$ increases, however, the measured EPR-Reid factors become too large to conclude on the presence of entanglement.
A common approach to certify entanglement in this regime, as used in~\cite{Edgar2012,PhysRevA.86.010101,Dąbrowski:17,HugoCertif}, is to estimate and subtract these events during post-processing. Figure~\ref{fig:EPR_Delta} shows that this subtraction enables violation of the EPR-Reid criterion even at higher $\Delta T$ for both cameras. However, as discussed in the introduction, this method relies on additional assumptions about the input state~\cite{PhysRevLett.120.203604}, which weakens the certification and introduces potential loopholes in practical applications.

\begin{figure}[htbp]
\centering\includegraphics[width=12cm]{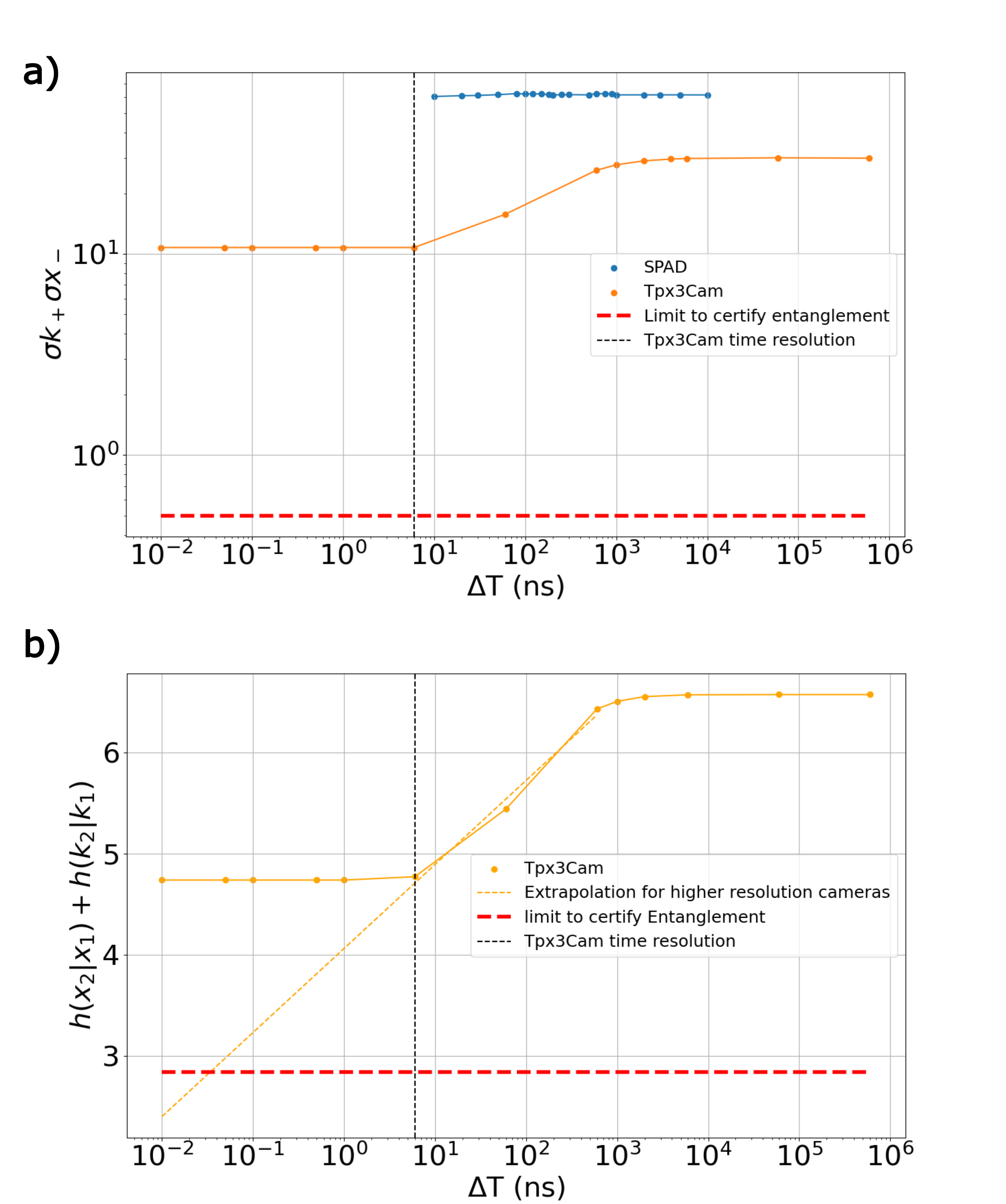}
\caption{a) EPR-Reid factor \(\Delta k_+ \Delta x_-\), estimated using the formal definition of variance for the correlation widths, as a function of the time resolution \(\Delta T\) with the Tpx3Cam. Each acquisition lasted 8 seconds for the Tpx3Cam and 650s for the SPAD camera.
Even at low \(\Delta T\), entanglement cannot be certified due to accidental coincidences, which increases the measured standard deviations.  
b) Sum of the conditional entropies in position and momentum, \(h(x_2|x_1) + h(k_2|k_1)\), as a function of \(\Delta T\) for the Tpx3Cam. For all \(\Delta T\) values, this criterion also fails to certify entanglement. In particular, for values below approximately $6$ns, reducing the time window $\Delta T$ does not improve the measured entropy, as the temporal resolution limit of the camera is reached. By extending the slope of the curve to these low values (dashed yellow line), one may infer that this criterion could allow entanglement certification with cameras reaching time resolutions on the order of 10–100ps}.
\label{fig:EPR_Sigma}
\end{figure}

In our work, violating the EPR-Reid inequality without accidental subtraction (Fig. 3) still relies on a key assumption: the Gaussian shape of the two-photon state. This allows us to extract the peak widths $\Delta x_-$ and $\Delta k_+$ by fitting the minus- and sum-coordinate projections with 2D Gaussians. However, as described in Ref.~\cite{reid_demonstration_1989,Cavalcanti_experimental_2009}, the correlation widths  are formally defined as variances i.e. $\Delta^2 x_- = \overline{\left(x_--\overline{x}_-\right)^2} $ and $\Delta^2 k_+ = \overline{\left(k_+-\overline{k}_+\right)^2}$.
Figure \ref{fig:EPR_Sigma}a shows the corresponding EPR-Reid factors $\Delta_{x-} \Delta_{k+}$ as a function of $\Delta T$. Interestingly, we observe that the inequality~\eqref{inequality} is never violated, preventing any conclusion about the presence of entanglement. 
Indeed, because the definition of variance gives the highest weight to the points further away from the center of the peak, this calculation method amplifies the impact of residual accidentals, leading to larger peak widths and higher EPR-Reid factors compared to Gaussian fitting.

Finally, the EPR-Reid criterion is a limiting case of a more general entropic criterion that can detect entanglement in situations where EPR-Reid fails~\cite{PhysRevLett.106.130402}. It uses the conditional entropies 
\begin{equation}
h(x_2|x_1) = \int dx_1 p(x_1) \left(-\int dx_2 p (x_2|x_1) \ln(p(x_2|x_1))\right)
\end{equation}
and 
\begin{equation}h(k_2|k_1) = \int dk_1 p(k_1) \left(-\int dk_2 p (k_2|k_1) \ln(p(k_2|k_1))\right),
\end{equation}
where $p(x_1)=\int p(x_1,x_2) dx_2$ and $p(x_2|x_1)$ is the probability of detecting the idler photon at $x_2$ conditioned on the detection of the signal photon at $x_1$. It enables to quantify how the measurement of one particle steers the state of the other. Entanglement is certified if  
\begin{equation}
h(x_2|x_1) + h(k_2|k_1) \leq \ln(2\pi e)
\end{equation}
for a given transverse spatial axis (\( \vec{x} \) or \( \vec{y} \)). The conditional entropies computed with the Tpx3Cam camera are shown in Figure~\ref{fig:EPR_Sigma}b. Without assuming a Gaussian two-photon state, this criterion also fails to certify entanglement, even for the shortest \( \Delta T \) values achievable with current state-of-the-art cameras. However, by extrapolating the evolution of the conditional entropies with $\Delta T$ (dashed yellow line in Figure~\ref{fig:EPR_Sigma}b), one may infer that entanglement could potentially be certified using this criterion - and under similar experimental conditions - with a camera reaching time resolutions on the order of 10–100 ps.

\section{Conclusion}
In this work, we experimentally studied how accidental coincidences affect camera-based entanglement certification protocols. We identified a regime - reachable with both Tpx3Cam camera and SPAD arrays - where accidentals become negligible to the point where entanglement can be certified by an EPR criterion without performing their subtraction. 
However, this certification still requires some hypotheses, most importantly the Gaussian fitting of the JPD projections to compute the correlation widths. Getting rid of this hypothesis with the EPR-Reid criterion protocol is currently prevented by the number of detected accidentals, even when using more permissive entropic criteria. 
To certify entanglement without relying on such a model, one possible solution would be to use cameras with improved temporal resolution — typically on the order of 10 to 100ps — which some SPAD camera models under development~\cite{mos_interleaved_2025} or future time-stamping cameras such as the Tpx4Cam \cite{Llopart_2022} could soon offer.
Another solution — which could be developed in parallel with the first — would involve designing a protocol that is more robust to accidental coincidences. This could include, for example, using a broader set of continuous-variable spatial bases~\cite{PhysRevA.94.012303}, or carefully selecting discrete bases defined by camera pixels to ensure proper mutual unbiasedness~\cite{PhysRevLett.120.040403}.

\begin{backmatter}
\bmsection{Funding} 
H.D. acknowledges funding from an ERC Starting Grant (grant no. SQIMIC-101039375).
\bmsection{Authors contributions} 
R.G., B.C. and C.V. designed the experiment and analyzed the data. R.G. performed the experiments. R. G. and H. D. conceived the original idea. A.N and P.S designed softwares to operate the Tpx3Cam camera. All authors discussed the results and contributed to the manuscript. H.D. supervised the project.
\bmsection{Disclosures} The authors declare no conflicts of interest.
\bmsection{Data availability} Data underlying the results presented in this paper are not publicly available at this time but may be obtained from the authors upon reasonable request.
\end{backmatter}

\bibliography{sample}

\newpage
\setcounter{figure}{0}
\setcounter{section}{0}
\setcounter{equation}{0}
\title{Supplementary Document : \\ Accidental coincidences in camera-based high-dimensional entanglement certification}

\author{Raphael Guitter \authormark{1,*}, Baptiste Courme \authormark{1,2}, Chloé Vernière \authormark{1}, Peter Svihra 
\authormark{3}, Andrei Nomerotski 
\authormark{3} and Hugo Defienne \authormark{1}}

\address{\authormark{1}Sorbonne Université, CNRS, Institut des NanoSciences de Paris, INSP, F-75005 Paris, France\\
\authormark{2}Laboratoire Kastler Brossel, ENS-Universite PSL, CNRS, Sorbonne Universite, College de France, 24 rue Lhomond, 75005 Paris, France\\}

\address{\authormark{3} Faculty of Nuclear Sciences and Physical Engineering,Czech Technical University, 160 00 Prague, Czech Republic}
\email{\authormark{*}raphael.guitter@insp.jussieu.fr} 

\section{Detailed experimental setups} \label{SupSec:Complete_Setups}
\begin{figure}[htbp]
\centering\includegraphics[width=12cm]{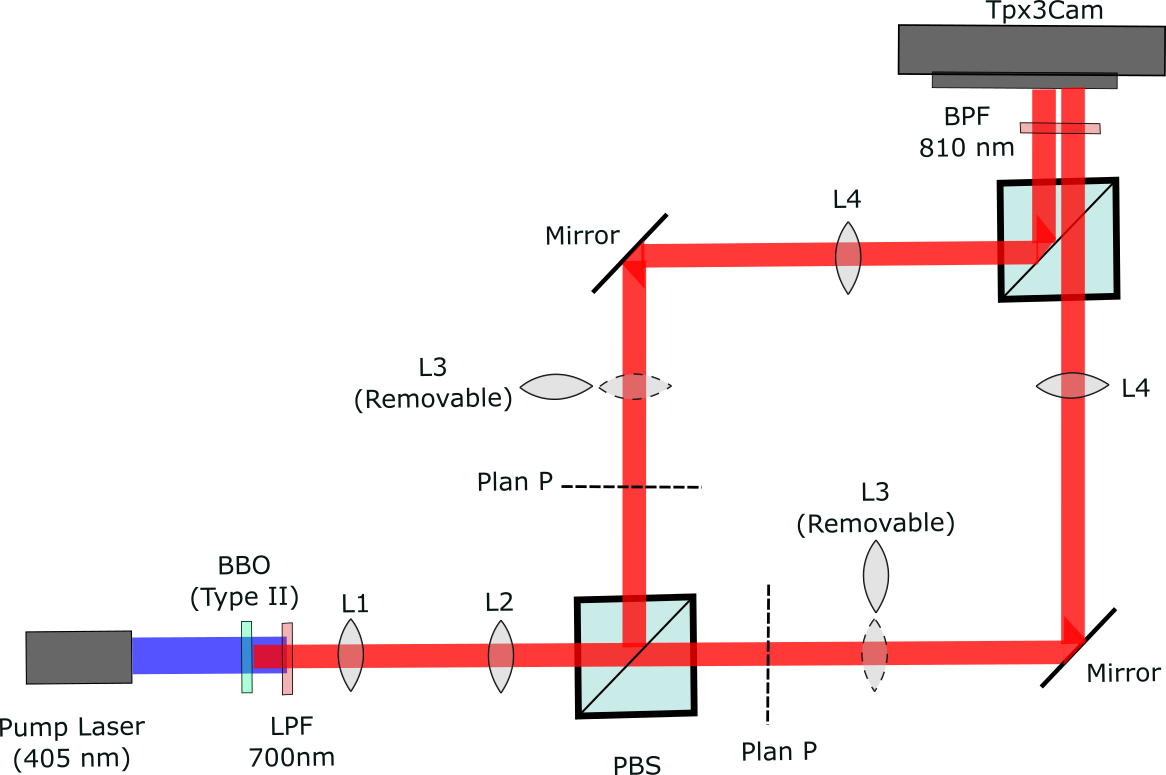}
\caption{Setup used to measure correlations in position and momentum with a Tpx3Cam. One can switch from imaging to standard imaging configuration by keeping or removing lens L3. Using Type II SPDC allows to separate the two photons of each pair using a Polarizing Beam Splitter (PBS) to send them on a separate area of the camera.
BBO: $\beta$-baryum borate nonlinear crystal, L:Lens, LPF: Long Pass Filter, PBS: Polarizing Beam Splitter, BPF: Band Pass Filter}
\label{fig:Setup_Tpx}
\end{figure}

\textbf{Tpx3Cam setup : } The complete setup is shown in Figure \ref{fig:Setup_Tpx}. The pump is a collimated continuous-wave laser at 405 nm from Oxxius with
an output power of 50 mW. The BBO crystal has dimensions $1\times 5 \times 5$ mm and is cut for type II SPDC at 405 nm. The crystal is slightly rotated around horizontal axis to ensure near-collinear phase matching of photons at the output (i.e. SPDC ring collapsed into a disk). A 700 nm-cut-off long-pass filter is used to block pump photons after the crystals. In the standard imaging configuration, the 4 f imaging system L1 - L2 in Figure 1.a of the manuscript is in reality composed of two successive 4 f systems with focal lengths f1=50mm, f2=150mm, f3=75mm, f4=150mm. The resulting magnification factor is equal to 6.
In the Fourier imaging configuration, the lens f3 is removed, resulting in a setup equivalent to the one shown in Fig 1.b) of the manuscript with an effective focal lens of 50mm. 
A band-pass filter at 810 $\pm$ 10 nm is placed just before the camera to only keep the degenerate photon pairs as well as filter out residual pump and ambient photons.
The camera is a Tpx3Cam camera from Amsterdam Scientific Instruments, combined with an image intensifier (Cricket, from Photonis, fitted with a 18mm dual-MCP Image Intensifier with a Hi-QE Red photocathode and P47 phosphor). 
The intensifier has a quantum efficiency of approximately
20$\%$  at 810nm. The camera has 256 $\times$ 256 pixels of 55 $\times$ 55 $\mu$m each. Each pixel of the camera operates and is read out independently with time resolution of 1.56ns \cite{NOMEROTSKI201926} and a dead time of about one microsecond. In practice, however, the timing dependence on the signal amplitude (timewalk effect) and small pixel to pixel time offsets reduce the effective temporal resolution to about 6 ns (FWHM) \cite{Vidyapin2023}. The losses in the experiment are mainly due to the quantum efficiency of the image intensifier, which leads to a total transmission efficiency of 0.2 \cite{Vidyapin2023}.

\begin{figure}[htbp]
\centering\includegraphics[width=12cm]{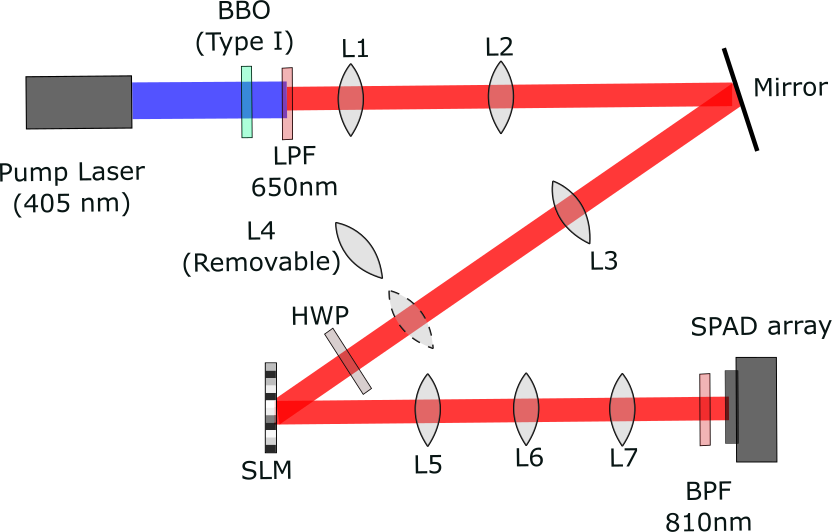}
\caption{Setup used to measure correlations in position and momentum with a SPAD array. One can switch from the imaging configuration to the Fourier imaging configuration by removing or keeping lens L4. The SLM is used as a grating acting only on the polarization at  45\textdegree \ compared to the one of the photons. As a results, it gives half of the photons a momentum offset, which translates to a position offset on the camera and allows to compute correlations between two different zones of the camera. BBO: $\beta$-baryum borate nonlinear crystal, L:Lens, LPF: Long Pass Filter, PBS: Polarizing Beam Splitter, BPF: Band Pass Filter, HWP: Half-Wave Plate}
\label{fig:Setup_SPAD}
\end{figure}

\textbf{SPAD array setup :}
This setup is presented in Figure \ref{fig:Setup_SPAD}.
The pump is a collimated continuous-wave laser at 405 nm  from Oxxius with an output power of 100 mW and a beam diameter of 0.8 ± 0.1 mm. BBO crystal has dimensions 0.5 $\times$ 5 $\times$ 5 mm and is cut for type I SPDC at 405 nm. A 650 nm-cut-off long-pass filter is used to block pump photons after the crystals.
In the Fourier imaging setup, the optical system is composed of 3 4f systems with focal lengths f1=50mm, f2=150mm, f3=100mm, f4=200mm, f5=150mm, f6=50mm. A last lens of focal length f7=150mm performs the optical Fourier transform onto the camera. Overall, the effective focal lens of this setup is 75mm. 
In the standart imaging configuration, the lens f4 is removed, which results in a standard imaging system of magnification 9. 
In both cases, the combination of a half-wave plate (HWP) rotated by $22.5$ degrees and  Spatial Light Modulator (SLM), displaying a phase ramp and placed between f4 and f5, is used to create a collinear beam splitter: since the SLM only acts on the horizontal polarization of the incoming light, half of the photons are directed to the $0^{th}$ order that is imaged on the left part of the camera, while the other half is directed to the $1^{th}$ order imaged on the right part of the camera. 
 A band-pass filter at 810 $\pm$ 10 nm is placed just before the camera to only keep the degenerate photon pairs as well as filter out residual pump and ambient photons.
 The camera is a SPAD SPC3 camera from Micro Photon Devices, with a photon detection efficiency around 5$\%$. The camera has 32 $\times$ 64 pixels of 150 $\times$ 150 $\mu$m each. Due to the readout electronics, the camera can take at most $10^5$ frames per second, which corresponds to a frame exposure time of 10 $\mu$s.

\section{Correlation peaks in the position basis}
\begin{figure}[htbp]
\centering\includegraphics[width=12cm]{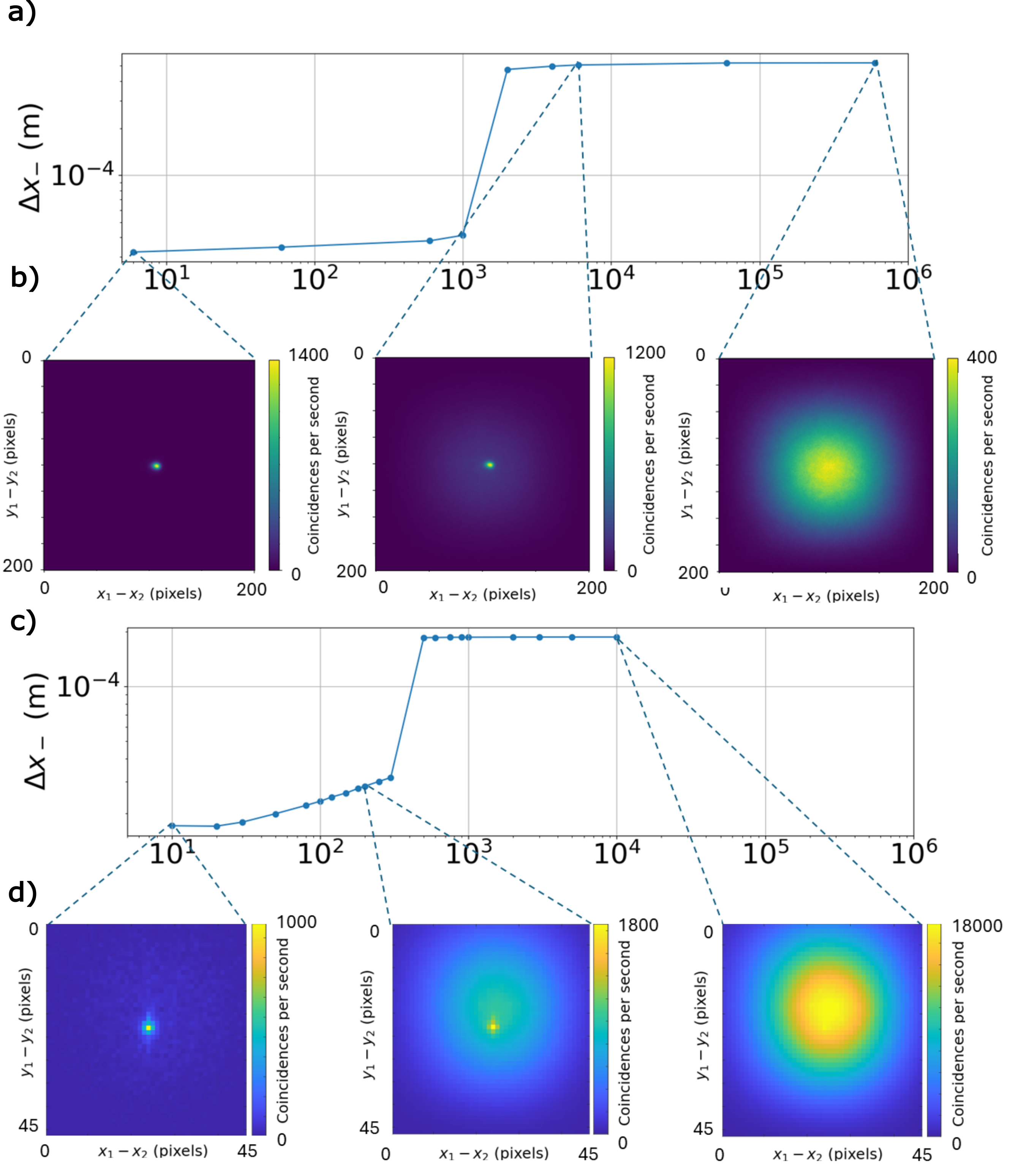}
\caption{a) Evolution of the position correlation peak width $\Delta x_-$ with the time window $\Delta T$ used to compute coincidences in the events detected by the Tpx3Cam time-stamping camera. b) Correlation images for different time resolutions. For $\Delta T$ of the order of a few ns, much lower that the average separation between two photon pairs (around 100 ns in our setup), only the true coincidences are recorded and the correlation peak remains narrow and $\Delta x_-$ small. As $\Delta T$ increases, the accidental coincidences become more prominent, widening the peak and increasing the value of $\Delta x_-$ by an order of magnitude. c)  Evolution of the position correlation peak width with the frame exposure time $\Delta T$ or the SPAD array camera. d) Correlation images for different time resolutions. The same phenomena can be observed as with the Tpx3Cam. }
\label{fig:Deltax}
\end{figure}

The effect of the time resolution of the coincidence counting on the measured correlation peak width $\Delta x_-$ in the position basis is presented is Figure \ref{fig:Deltax}. The same behavior that is observed in the momentum basis can also be observed in the position basis. 

\section{Gaussian Model}
In the Gaussian approximation, the two-photons wave function $\Psi $ is approximated by a double Gaussian~\cite{fedorov_gaussian_2009} 
\begin{equation}
    \Psi(x_1,y_1,x_2,y_2) = A \exp\left(-\frac{(x_1-x_2)^2 + (y_1 - y_2)^2}{4 \Delta x_-^2}\right) \exp\left(-\Delta k_+^2 \frac{(x_1+x_2)^2 + (y_1 + y_2)^2}{4}\right).
\end{equation}
where $\Delta x_-$ and $\Delta k_+$ coincide with the variances used in the EPR-Reid criterion (Eq 1 in Main Text).
In that case, the minus-coordinate projection of the JPD in the position basis is also Gaussian and takes the form 
\begin{equation}
    \Gamma(x_1 - x_2, y_1 - y_2) \propto \exp\left(-\frac{(x_1-x_2)^2 + (y_1 - y_2)^2}{2 \Delta x_-^2}\right)
\end{equation}
and the sum-coordinate projection in the momentum basis takes the form 
\begin{equation}
    \Gamma(k_{x1} + k_{x2}, k_{y1} + k_{y2}) \propto \exp\left(-\frac{(k_{x1} + k_{x2})^2 + (k_{y1} + k_{y2})^2}{2 \Delta k_+^2}\right)
\end{equation}

which allows to directly retrieve the quantities $\Delta x_-$ and $\Delta k_+$ by a Gaussian fit of these projections. 

\section{Uncertainties}
The values of $\Delta k_+$ and $\Delta x_-$ presented in this work are obtained by a 2D Gaussian fit of the sum- and minus-coordinate projections with a model
\begin{equation}
f(x,y) = Ae^{-\frac{(x-x_0)^2 + (y-y_0)^2}{2\Delta^2}}
\end{equation}
where $\Delta$ is then used to determine $\Delta k_+$ or $\Delta x_-$. The fitted projections show noise that can be characterized by measuring the standard deviation $\Sigma$ of this noise in the area outside of the correlation peak (measured over 25 $\times 50$ pixels for the Tpx3Cam Data and over 5 $\times$ 10 pixels for the SPAD Data). This $\Sigma$ is linked to the uncertainty $u_\Delta$ of the measured $\Delta$ by computing the gradient of f at position $\sqrt{(x-x_0)^2 + (y-y_0)^2} = \Delta$ to obtain 
\begin{equation}
    u_\Delta = \frac{\sqrt{e} \Delta}{a} \Sigma
\end{equation}
The specific uncertainties of $\Delta k_+$ and $\Delta x_-$ are then computed by taking into account the magnification and effective focal length of the imaging setup. 
\end{document}